\documentclass[12pt]{article}
\usepackage{amsmath}
\usepackage{amsthm}
\usepackage{amssymb}
\usepackage{mathrsfs}
\usepackage[usenames]{color}
\usepackage[latin5]{inputenc}
\usepackage{cite}
\usepackage{pdfsync}
\usepackage{enumitem}
\setlist{parsep=0pt,itemindent=0pt}
\definecolor{webgreen}{rgb}{0,0.5,0}

\theoremstyle{plain}

\theoremstyle{definition}

\numberwithin{equation}{section}
\numberwithin{thm}{section}
\numberwithin{lemma}{section}
\numberwithin{prop}{section}
\numberwithin{cor}{section}
\numberwithin{rmk}{section}
\numberwithin{defn}{section}
\numberwithin{exa}{section}

\newcommand{\gen}[1]{\partial_{#1}}

\newcommand{\curl}[1]{ \left\{#1\right\} }
\newcommand{\lie}{\mathfrak g}

\definecolor{darkolivegreen}{rgb}{0.333333, 0.419608, 0.1843140}

\begin{document}
\pagenumbering{arabic}
\clearpage
\thispagestyle{empty}

\title{Calculation of the propagator of Schr\"odinger's equation on $(0,
  \infty)$ with the potential $kx^{-2} + \omega^2x^2$ by Lie symmetry group method}

\author{
F.~G\"ung\"{o}r\\ \small
Department of Mathematics, Faculty of Science and Letters,\\ \small Istanbul Technical University, 34469 Istanbul, Turkey \thanks{e-mail: gungorf@itu.edu.tr}}

\date{}

\maketitle



\begin{abstract}
The propagators (fundamental solutions) of the heat and Schr\"odinger's equations on the half-line with a combined harmonic oscillator and inverse-square potential calculated in the recent paper
[{\em  J. Math. Phys.} {\bf 59}, 051507 (2018)]
using Laplace's method are demonstrated to   be obtainable alternatively within the framework of symmetry group methods discussed in a series of two papers in the same journal.
\end{abstract}

\vspace{1,5mm}


\section{Introduction}
Very recently, in Ref. \cite{OrtnerWagner2018},  Green function of  the operator
\begin{equation}\label{parab-op}
  P(\partial)=\gen t+\frac{1}{2}(-\partial_x^2+V(x))=\gen t-\frac{1}{2}\partial_x^2+\frac{k}{2}x^{-2}+\frac{1}{2}\omega^2 x^2, \qquad k\geq -\frac{1}{4}, \quad \omega\geq 0
\end{equation}
has been calculated in the half-space $x>0$ and hence from it  that of Schr\"odinger's equation with the potential
$$V(x)=\frac{1}{2}\partial_x^2+\frac{k}{2}x^{-2}+\frac{1}{2}\omega^2 x^2,  \quad k\geq \frac{3}{4}$$ is deduced using Laplace's method.

In another recent  paper  \cite{Guengoer2018a} with a reliance on the results of \cite{Guengoer2018}, we studied calculation of fundamental solutions of variable coefficient linear parabolic equations allowing sufficiently enough symmetry groups.

The purpose of this paper is to make use of methods from \cite{Guengoer2018, Guengoer2018a} as an alternative natural approach to  the one pursued in \cite{OrtnerWagner2018} to derive fundamental solutions for the parabolic operator \eqref{parab-op} and its Schr\"odinger variant.

\section{Two equivalent definitions of fundamental solution}
Fundamental solution  (the names Green function, source solution, propagator, heat kernel  are also used interchangeably) for the initial value problem  $u_t-Lu=0$, $t>0$,  $u(x,0)=f(x)$, $x\in \mathbb{R}^n$,  where $L=\sum_{|\alpha|\leq m}a_\alpha(x)D^{\alpha}$, $\alpha\in \mathbb{N}^n$ with the property $a_\alpha\in C^{|\alpha|}(\mathbb{R}^n)$ is a linear differential operator of order $m$,  can be defined in two  ways. One way is to define a distribution function $E_{\xi}(t,x)\in\mathcal{D'}(\mathbb{R}^n)$, $t\geq 0$ which solves the following initial value problem
\begin{equation}\label{def-1}
  (\partial_t-L)E_{\xi}(t,x)=0,  \quad E_{\xi}(0,x)=\delta(x-\xi), \quad x,\xi\in \mathbb{R}^n
\end{equation}
This is the definition used in Ref. \cite{Guengoer2018a}.

Another definition is the free space (also called the causal) fundamental solution, which satisfies
\begin{equation}\label{def-2}
  (\partial_t-L)\tilde{E}_{\xi}(t,x)=\delta(t)\delta(x-\xi),  \quad \text{in} \; \; \mathcal{D}'(\mathbb{R}^{n+1}).
\end{equation}
This definition is used in Ref. \cite{OrtnerWagner2018}. It can be proved that the fundamental solution $E_{\xi}(t,x)$ as defined in \eqref{def-1} coincides with the free-space one when one extends $E_{\xi}(t,x)$ by zero to $t\leq 0$
\begin{equation}\label{extension}
E'_{\xi}(t,x)=H(t)E_{\xi}(t,x)=
\begin{cases}
  E_{\xi}(t,x), & \mbox{for} \quad t>0 \\
  0, & \mbox{for} \quad t < 0.
\end{cases}
\end{equation}
The proof that for distributions $E'_{0}(t,x)$ satisfies $(\partial_t-L)E'_{0}(t,x)=\delta(t,x)=\delta(t)\delta(x)$  can be found in the book \cite{McOwen1996} for the heat operator, i.e. when $L=\triangle$ is the Laplace operator.

Consequently, all the fundamental solutions constructed here can be easily made to agree with those of Ref. \cite{OrtnerWagner2018} by their extension by zero to $t\leq 0$ just by multiplying them by the Heaviside function $H(t)$ (or $Y(t)$ by the notation of \cite{OrtnerWagner2018}).

\section{Calculation of fundamental solution (Green function)}

We first consider the operator
\begin{equation}\label{op-1}
  L=P(\partial)=\gen t+\frac{1}{2}(-\partial_x^2+V(x))=\gen t-\frac{1}{2}\partial_x^2+\frac{k}{2}x^{-2}+\frac{1}{2}\omega^2 x^2, \qquad k\geq -\frac{1}{4}, \quad \omega\geq 0
\end{equation}
and construct the Green function $E_{\xi}(t,x)\in \mathcal{D}'(\mathbb{R}_{+})$, $\xi>0$ in the half-line in the sense of first definition \eqref{def-1} for the heat equation with the potential $V(x)$
\begin{equation}\label{op-action-1}
  Lu=P(\partial)u=u_t-\frac{1}{2}u_{xx}+\left(\frac{k}{2}x^{-2}+\frac{1}{2}\omega^2 x^2\right)u=0.
\end{equation}
We can get rid of the factors of 1/2 by  scaling $t$, $t\to t/2$. We hence consider
\begin{equation}\label{heat-eq}
  u_t=u_{xx}-(kx^{-2}+\omega^2 x^2)u.
\end{equation}

A brief summary of the symmetry approach is as follows.
We look for a transformation that leaves invariant the initial value problem \eqref{def-1} on the half-line $x>0$. Such transformation is realized as a symmetry group of the Lie algebra spanned by the vector fields of the form
\begin{equation}\label{vf}
  \mathbf{v}=\tau(t)\gen t+\chi(t,x)\gen x+\phi(t,x)u\gen u,
\end{equation}
satisfying the conditions
\begin{equation}\label{conds}
  \tau(0)=0,  \quad \chi(0,\xi)=0,  \quad \phi(0,\xi)+\chi_x(0,\xi)=0.
\end{equation}
Fundamental solutions then arise as solutions invariant under this symmetry group.
For more details of this and another related method  the reader is directed to Ref. \cite{Guengoer2018a}.

We know  from the results of \cite{Guengoer2018} that this equation admits a four-dimensional Lie point symmetry algebra $\lie$, excluding the obvious infinite-dimensional one, because $k\ne 0$ (otherwise 6-dimensional). From the formula (3.28) of Ref. \cite{Guengoer2018} (correcting  sign errors in the vector field $\mathbf{v}_3$; all signs appearing in the exponentials should be negative) with $a=1$, $b=0$, $c=-V(x)$, $I(x)=x$, $J(x)=0$, $c_2=-\omega^2$, $c_0=0$ we find the following basis of the symmetry algebra
\begin{equation}\label{4-dim-L}
\begin{split}
&\mathbf{v}_1=\gen t,\\
&\mathbf{v}_2=\cosh (4\omega t)\gen t+2\omega \sinh (4\omega t)x\gen x-[\omega \sinh (4\omega t)+2\omega^2 \cosh (4\omega t)x^2] u\gen u,\\
&\mathbf{v}_3=\sinh (4\omega t)\gen t+2\omega \cosh (4\omega t)x\gen x-[\omega \cosh (4\omega t)+2\omega^2 \sinh (4\omega t)x^2] u\gen u,\\
&\mathbf{v}_4=u\gen u
\end{split}
\end{equation}
with non-zero commutation relations
$$[\mathbf{v}_1,\mathbf{v}_2]=4\omega \mathbf{v}_3,  \quad [\mathbf{v}_1,\mathbf{v}_3]=4\omega \mathbf{v}_2,  \quad [\mathbf{v}_2,\mathbf{v}_3]=4\omega \mathbf{v}_1.$$
The general symmetry vector field $\mathbf{v}=\sum_{i=1}^4 k_i \mathbf{v}_i$  satisfying the  initial-boundary conditions \eqref{conds} is given by
\begin{equation}\label{gen-vf-1}
  \mathbf{v}=\sinh^2 (2\omega t)\gen t+2\omega \sinh (4\omega t)x\gen x+\omega[2\omega \xi^2-2\omega \cosh (4\omega t) x^2-\sinh (4\omega t)]u\gen u.
\end{equation}
We find the invariants of $\mathbf{v}$ by solving the equation $\mathbf{v}(\varphi)=0$ by the method of characteristics  as
\begin{equation}\label{inv-1}
  \eta=\frac{x}{\sinh (2\omega t)},  \quad \zeta=u^{-1}{\sinh (2\omega t)}^{-1/2}\exp\left[-\frac{\omega(x^2+\xi^2)}{2\tanh(2\omega t)}\right].
\end{equation}
Fundamental solution will be sought as a group-invariant solution in the form
\begin{equation}\label{G-inv-1}
  u={\sinh (2\omega t)}^{-1/2}\exp\left[-\frac{\omega(x^2+\xi^2)}{2\tanh(2\omega t)}\right]F(\eta).
\end{equation}
Substitution of \eqref{G-inv-1} into \eqref{heat-eq} provides the second order ordinary differential equation (ODE)
\begin{equation}\label{ODE-1}
  \eta^2 F''(\eta)-(\omega^2\xi^2 \eta^2+k)F=0.
\end{equation}
The modified Bessel equation of  with index $\nu$ in normal form is
$$\eta^2 F''(\eta)-(\omega^2\xi^2 \eta^2-\frac{1}{4}+\nu^2)F=0.$$
Hence, the solution of \eqref{ODE-1} bounded near zero can  be written as
$$F=\sqrt{\eta}I_{\nu}(\omega \xi \eta),  \quad \nu=\sqrt{k+\frac{1}{4}}\geq 0,$$
where $I_{\nu}$ is the modified Bessel function of the first kind with index $\nu$.
Finally, the fundamental solution $E_{\xi}(t,x)$ of \eqref{op-action-1} is determined up to a multiplicative constant $c_0$ (after the replacement $t\to t/2$)
\begin{equation}\label{Green-1}
  E_{\xi}(t,x)=c_0\frac{\sqrt{x}}{\sinh \omega t}\exp\left[-\frac{\omega(x^2+\xi^2)}{2\tanh(\omega t)}\right]I_{\nu}\left(\frac{\omega \xi x}{\sinh \omega t}\right).
\end{equation}
The constant $c_0$ should be found from the normalization condition $$\lim_{t\to 0^{+}}\int _{0}^{\infty}E_{\xi}(t,x)dx=1.$$


Green function for the special potential when $\omega=0$ has already be obtained in Ref. \cite{Guengoer2018a} (page 6) using  methods within the symmetry context (with $k=-\mu$ in notation of \cite{Guengoer2018a}). We reproduce it here for the purpose of reference (the replacement $t\to t/2$ is done)
$$E_{\xi}(t,x)=\frac{\sqrt{\xi x}}{t}\exp{\left[-\frac{x^2+y^2}{2t}\right]}I_{\nu}\left(\frac{xy}{t}\right), \quad \nu=\sqrt{\frac{1}{4}+k}.$$

We now turn our attention to the construction of Green function for the one-dimensional Schr\"odinger equation with potential $V(x)$
\begin{equation}\label{Schr-eq}
iu_t+u_{xx}=(k x^{-2}+\omega^2 x^2)u,
\end{equation}
where $u:\mathbb{R}_{+}^{2}=\curl{(t,x)\in\mathbb{R}^2:x>0}\to  \mathbb{C}$ is a complex-valued function.
From \cite{Guengoer2018}, we recall that the symmetry algebra for an arbitrary potential $V(t,x)$ (with some sign corrections) is represented by
$$\mathbf{v}=\tau(t)\gen t+\chi(t,x)\gen x+ i \phi(t,x)u\gen u,$$
where
$$\chi(t,x)=\frac{1}{2}\dot{\tau}x+\rho(t),  \quad \phi(t,x)=\frac{\ddot{\tau}}{8}x^2+\frac{\dot{\rho}}{2}x+\sigma(t)+i\left(\frac{\dot{\tau}}{4}+b\right)$$
and $V$ satisfies the determining equation
\begin{equation}\label{det}
  \tau V_t+\chi V_x+\dot{\tau}V+\frac{\dddot{\tau}}{8}x^2+\frac{\ddot{\rho}}{2}x+\dot{\sigma}(t)=0.
\end{equation}
Here $b$ is a  constant, $\tau(t)$, $\rho(t)$ and $\sigma(t)$ are arbitrary real functions.
For the given potential $V=V(x)=k x^{-2}+\omega^2 x^2$, Eq. \eqref{det} is easily solved for these functions, and it turns out that  equation \eqref{Schr-eq} admits a five-dimensional symmetry algebra. A basis for its elements is given by
\begin{equation}\label{4-dim-L-Schr}
\begin{split}
&\mathbf{v}_1=\gen t,\\
&\mathbf{v}_2=-\frac{\cos 4\omega t}{4\omega}\gen t+\frac{\cos 4\omega t}{2}x\gen x+\frac{1}{4}\left[2i \omega \cos (4\omega t) x^2-\sin 4\omega t\right]u\gen u,\\
&\mathbf{v}_3=\frac{\sin 4\omega t}{4\omega}\gen t+\frac{\sin 4\omega t}{2}x\gen x-\frac{1}{4}\left[2i \omega \sin (4\omega t)x^2+\cos 4\omega t\right]u\gen u,\\
&\mathbf{v}_4=u\gen u+u^{*}\gen {u^{*}},   \quad \mathbf{v}_5=i(u\gen u-u^{*}\gen {u^{*}}),
\end{split}
\end{equation}
where $*$ denotes the complex conjugation and in $\mathbf{v}_2$ and $\mathbf{v}_3$ complex conjugated coefficients of $u$-component are omitted.
Again, the most general symmetry element leaving the initial condition $\lim _{t\to 0^{+}}E_{\xi}(t,x)=\delta(x-\xi)$ invariant has the form
\begin{equation}\label{gen-vf-2}
 \mathbf{v}=2\sin^2 (2\omega t) \gen t+2\omega\sin (4\omega t)x\gen x+i\omega\left[-2\omega \xi^2+2\omega \cos (4\omega t)x^2+i \sin (4\omega t)\right]u\gen u.
\end{equation}
Invariants are $\eta=x/\sin(2\omega t)$ and $\zeta=u^{-1}(\sin 2\omega t)^{-1/2}\exp[2^{-1}i\omega(x^2+\omega^2)\cot (2\omega t)]$. Green function will be of the form
\begin{equation}\label{G-inv-2}
  u=(\sin 2\omega t)^{-1/2}\exp\left[\frac{i\omega(x^2+\omega^2)}{2\tan(2\omega t)}\right]F(\eta).
\end{equation}
When substituted into \eqref{Schr-eq} we find that $F$ should satisfy the ODE
\begin{equation}\label{ODE-2}
 \eta^2 F''(\eta)+(\omega^2\xi^2 \eta^2-k)F(\eta)=0.
\end{equation}
which is the normal form of the Bessel equation with index $\nu=\sqrt{\frac{1}{4}+k}$. Therefore, the Green function of Eq. \eqref{Schr-eq}, up to a nonzero normalization constant, is given by \begin{equation}\label{Green-2}
  E_{\xi}(t,x)=c_0\frac{\sqrt{x}}{\sin \omega t}\exp\left[\frac{i\omega(x^2+\xi^2)}{2\tan(\omega t)}\right]J_{\nu}\left(\frac{\omega \xi x}{\sin \omega t}\right).
\end{equation}

The calculation of the Green function of \eqref{Schr-eq} for $\omega=0$ is slightly different in which the relevant symmetry vector field now becomes a projective type
\begin{equation}\label{gen-vf-3}
  \mathbf{v}=t^2\gen t+xt\gen x+\frac{1}{4}[i(x^2-\xi^2)-2t]u\gen u.
\end{equation}
In this situation, solution ansatz (using invariants of $\mathbf{v}$) will be in the form
\begin{equation}\label{G-inv-3}
  u=\frac{1}{\sqrt{t}}\exp[\frac{i(x^2+\xi^2)}{4t}]F(\eta), \quad \eta=\frac{x}{t},
\end{equation}
where the function $F$ satisfies
\begin{equation}\label{ODE-3}
  \eta^2F''(\eta)+\left(\frac{\xi^2}{4}\eta^2-k\right)F(\eta)=0
\end{equation}
with the appropriate solution $F=\sqrt{\eta}J_\nu(\frac{\xi \eta}{2})$, $\nu=\sqrt{\frac{1}{4}+k}$. Finally, the Green function (up to a normalization constant) is given by
\begin{equation}\label{Green-3}
  E_{\xi}(t,x)=c_0\frac{\sqrt{x}}{t}\exp[\frac{i(x^2+\xi^2)}{4t}]J_\nu(\frac{\xi x}{t}).
\end{equation}
$E_{\xi}(t,x)$ given in \eqref{Green-2} and \eqref{Green-3}  is smooth if $x\ne 0$, $(t,x)\ne (0,\xi)$,  namely of class $C^{\infty}(\mathbb{R}\times \mathbb{R}^n\setminus(0,\xi))$ (See \cite{OrtnerWagner2018} for their remarkable properties).


\end{document}